# Double layer in ionic liquids: capacitance vs. temperature from atomistic simulations

Heigo Ers[a], Iuliia V. Voroshylova[b], Piret Pikma[a], Vladislav B. Ivaništšev[a,c*]

[a] Institute of Chemistry, University of Tartu, Ravila 14a, Tartu 50411, Estonia

[b] LAQV@REQUIMTE, Faculdade de Ciências, Universidade do Porto, Departamento de Química e Bioquímica, Rua do Campo Alegre, 4169-007 Porto, Portugal

[c] Department of Chemistry, Center for High Entropy Alloy Catalysis, University of Copenhagen (UCPH), 2100 Copenhagen Ø, Denmark

Corresponding author email: vliv@chem.ku.dk

**Keywords**: graphene, differential capacitance, temperature dependence, interfacial structure, molecular dynamics, electrical double layer, 1-ethyl-3-methylimidazolium tetrafluoroborate

## Abstract

In this study, we investigated the graphene–ionic liquid (EMImBF$_4$) interface to clarify the effects of ambient temperature and potential on the differential capacitance. We complemented molecular dynamics simulations with density functional theory calculations to unravel the electrolyte and electrode contributions to the differential capacitance. As a result, we show: (1) the relation of characteristic saddle points of the capacitance–potential curve to the structural changes; (2) the smearing effect of temperature on the local structure and, consequently, on the capacitance; (3) rationalization of these observations with the interfacial bilayer model; and, finally, (4) how quantum capacitance correction dampens the influence of temperature and improves the agreement with the experimental data. These insights are of fundamental and practical importance for the application of similar interfaces in electrochemical energy storage and transformation devices, such as capacitors and actuators.

# 1. Introduction

Ionic liquids (ILs) are solvent-free electrolytes with exceptional properties. They are characterized by high electrochemical stability and low vapor pressure, making them attractive for use in fuel cells, solar panels, supercapacitors, and electric actuators [1–6]. The interfacial properties of ILs are also of fundamental interest, as their electrical double layer (EDL) differs significantly from the one formed in aqueous and organic electrolytes. That is due to the high concentration of ions, which leads to the layered IL structure and overscreening phenomenon [7,8]. For these reasons, the EDL in IL is actively studied theoretically [9–13], experimentally [14–17], as well as using computational chemistry methods such as molecular dynamics (MD) [18–23] and density functional theory (DFT) [24–26] with increasingly complex models [27–29]. Despite numerous studies, multiple open questions about ILs interfacial properties remain. One of such questions is the nature of differential capacitance ($C$) dependence on temperature ($T$).

The $T$-dependence of the EDL properties of ILs was previously studied both experimentally and computationally. Evaluating $C$ is of great interest here because it allows one to compare the derivative of $dC/dT$ from experimental and computational studies. The existing argumentations support both negative and positive $dC/dT$ derivatives. The negative one is consistent with the mean-field theory, while the positive one ("anomalous") is revealed in theory when accounting for the ion–ion correlation [30–32]. Several experimental works [33–37] investigating the interfaces of ILs consisting of $BF_4^-$, $PF_6^-$, TFI$^-$ or TFSI$^-$ anion and imidazolium (Im) or pyridinium (Pyr) ion-based cation with alkyl chains (M = methyl, E = ethyl, P = propyl, B = butyl), mostly with gold, platinum, or glassy carbon electrodes, showed that the increase in $T$ causes an increase in capacitance ($dC/dT > 0$). Silva *et al.* [34], Siinor *et al.* [33], and Lockett *et al.* [37] justified $dC/dT > 0$ by the thermal dissociation of ion associates, which allows ions to compensate for the charge of the electrode more effectively and thereby reduce the width of the charge layer in the electrolyte. Ivaništšev *et al.* [35] associated $dC/dT > 0$ with the mixing of ion layers when increasing $T$.

In contrast to the studies mentioned above, Drüschel *et al.* [38] showed that the capacitance decreases with increasing $T$ ($dC/dT < 0$) in an experimental study of the gold–BMPyrFAP interface. They associated the effect with two capacitive processes with different $T$-dependent relaxation times. Several computational works also revealed $dC/dT < 0$. Vatamanu *et al.* [39] studied the interface between PMPyrFSI and graphite with MD simulations. They reasoned $dC/dT < 0$ with the diminishing of the ordered layered structure due to the $T$ increase, which lowers the charge separation in the electrolyte. In the case of rough electrode surfaces, they also pointed out that ions can be in the surface cavities at low $T$. As the $T$ increases, the placement of ions in the cavities is hindered due to thermal motion, which increases the separation of the electrode and ions ($d$) and thus reduces the capacitance (assuming $C \sim d^{-1}$). Chen *et al.* [40] observed in MD simulations the $C(T)$ dependence only in the vicinity of the potential of zero charge (PZC). Liu *et al.* [41] showed in MD simulations of the graphite–BMImPF$_6$ interface that in general $dC/dT < 0$, yet in a narrow potential ($U$) range, capacitance increases. They associated $dC/dT < 0$ with the attenuation of BMIm$^+$ cation adsorption on the graphite surface, allowing the anions to screen the surface charge more effectively. With classical DFT simulations, Shen *et al.* [42] also demonstrated that the effect of $T$ on $C(U)$ is $U$-dependent and can switch from negative to positive. The $dC/dT > 0$ was also assumed by Kislenko *et al.* [43] by evaluating potential drops in MD simulations of the graphite–BMImPF$_6$ interface. To conclude, the understanding of the dependence and its underlying causes remains unclear, as different experimental and computational studies described both negative and positive $dC/dT$ dependences within the studied $U$ ranges.

In this work, we modeled the interface between 1-ethyl-3-methylimidazolium tetrafluoroborate (EMImBF$_4$) and graphene (Gr) using MD and DFT. We focused on the capacitance and structure dependence on the potential and temperature. The MD simulations allowed us to characterize the changes in the structure of the IL near the Gr surface, while the DFT calculations accounted for the semimetallic nature of Gr.

## 2. Methods

### 2.1. The studied model and computational method

To simulate the interface between Gr and EMImBF$_4$ at different $T$ and $U$ values, a model consisting of two rigid Gr sheets with electrodes spaced 9.5 nm apart was constructed. Each Gr sheet consisted of 336 carbon atoms and had an area of 2.98×2.95 nm². The space between the electrodes was filled with 288 EMIm$^+$ and 288 BF$_4^-$ ions, shown in Fig. 1a, using the Packmol software package [44]. The ions represented an IL with a density close to the experimental data [45].

MD simulations with the constructed model were run stepwise. First, energy minimization was performed using the gradient descent method while maintaining $T$ at 450 K. The energy minimization was followed by two equilibration steps with durations of 0.1 ns (time step 0.5 fs) and 1 ns (time step 1.0 fs), both at 450 K. Followingly, the system was then annealed at 1000, 900, or 800 K for 0.5 ns with a time step of 2.0 fs. Next, to represent the charging of the electrodes with opposite but equal charges, an electric field ($E$) was applied to the systems in a direction perpendicular to the Gr electrode surface. The $E$ values corresponded to the surface charge density (σ) in the range |0–70| μC/cm² and were calculated as $E = \sigma / (\epsilon\epsilon_0)$ using the value of 1.6 as relative permittivity ($\epsilon$). After that, the system was cooled down to the $T$ of 300, 350, 400, or 450 K. The charging and cooling steps both lasted 10 ns (timestep 1.0 fs). As a result, three replicas were obtained for each given $E$ and $T$. As a final step, to collect the data for further analysis, an additional simulation of 10 ns (1.0 fs timestep) was performed with each of the obtained systems, using a pre-set $T$ and $E$, defined in previous steps. Snapshots of the simulations were stored every 10 ps.

The NaRIBaS framework was used to prepare each described simulation step [46]. All MD simulations were conducted on the Rocket cluster of HPC Center of the University of Tartu [47] using Gromacs versions 5.1.4 and 2019.5 [48,49]. Simulations were performed in an *NVT* ensemble using the OPLS-AA force field developed by Lopes *et al.* [50], periodic boundary conditions in directions parallel to the surface plane, and a velocity rescale thermostat [51].

In addition, to estimate the density of states (DOS) for an isolated Gr sheet, DFT calculations were performed using the GPAW 21.6.0 software [52,53]. The calculations were performed using the plane-wave method with a cutoff of 800 eV, Perdew–Burke–Ernzerhof (PBE) exchange-correlation functional [54], and a Gr sheet, consisting of 8 carbon atoms and having an area of 0.21 nm², generated using the ASE Python library [55]. For the Brillouin zone sampling, a 40×40×1 Monkhorst–Pack $k$-point grid was used [56]. The periodic boundary conditions were applied in all directions while adding 1 nm of vacuum perpendicular to the Gr surface. All other parameters used were default settings for the GPAW software.

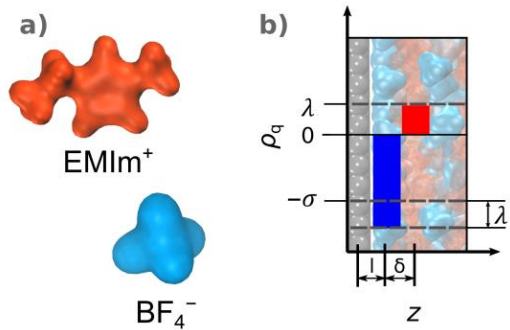

Figure 1. a) The vdW iso-surfaces of simulated IL ions. b) Profiles of the positively charged Gr–EMImBF$_4$ interface, illustrating the variables of the interfacial bilayer model. Cations and anions are shown with red and blue colors, respectively.

## 2.2. Evaluation of Potential Drop and Electrolyte Capacitance

To evaluate the contributions of Gr and IL to the potential drop and interfacial capacitance, we divided the interface into the electrolyte and electrode components. The electric potential ($\varphi(z)$) of the IL due to non-uniform distribution of ions at distance $z$ from the electrode surface was found for each replica of a simulated $T$ and $\sigma$ by integrating the Poisson equation:

$$\varphi(z) = -\frac{z\sigma}{\epsilon\epsilon_0} - \frac{1}{\epsilon\epsilon_0}\int_0^z (z - z')\rho_q(z')dz', \quad (1)$$

where $\sigma$ is the surface charge density, $\rho_q(z')$ IL charge density at a distance $z'$ from the electrode averaged over trajectory snapshots. $\epsilon_0$ is the permittivity of vacuum, and $\epsilon$ is the high-frequency permittivity describing only the electronic polarization.

The potential drop ($\varphi_{IL}$) in the IL electrolyte due to non-uniform distribution of ions at the electrode–IL interface was estimated as:

$$\varphi_{IL} = \varphi_{electrode} - \varphi_{electrolyte}, \quad (2)$$

where $\varphi_{electrode}$ and $\varphi_{electrolyte}$ are electric potentials at the electrode surface and in the IL bulk, respectively. In the following analyses, instead of the $\varphi_{IL}$, relative potential ($U$) was used:

$$U = \bar{\varphi}_{IL} - \bar{\varphi}_{PZC}, \quad (3)$$

where $\bar{\varphi}_{IL}$ is the mean average potential drop of all simulated replicas and $\bar{\varphi}_{PZC}$ is the mean average potential of zero charge (PZC).

The electrolyte capacitance ($C_{IL}$) was evaluated as:

$$C_{IL} = \frac{d\sigma}{dU}. \quad (4)$$

To smooth the interpolated $\sigma$–$U$ dependence, the convolution smoothing with a Hamming window function with a window size of 12 µC cm$^{-2}$ was used. The positive and negative aspects of this smoothing method are discussed in Ref. [57].

For the estimation of electrode quantum capacitance ($C_Q$), the $\sigma$ at Gr electric potential ($\varphi_{Gr}$), was calculated as:

$$\sigma = e \int_{-\infty}^{\infty} D(\varepsilon)[f(\varepsilon, T) - f(\varepsilon - e\varphi_{Gr}, T)]d\varepsilon, \quad (5)$$

where $D(\varepsilon)$ is the Gr DOS, and $f(\varepsilon, T)$ is the Fermi–Dirac distribution function at the given $T$ and electronic energy level $\varepsilon$, shifted relative to Gr Fermi level ($E_F$). The value of $\varphi_{Gr}$ was estimated from DOS by $\varphi_{Gr} = (\varepsilon - E_F)/e$. Then, $C_Q$ was calculated as:

$$C_Q = \frac{d\sigma}{d\varphi_{Gr}}, \quad (6)$$

using convolution smoothing with a smaller window size of 1.25 µC cm$^{-2}$ to avoid oversmoothing the capacitance minima at $\varphi_{Gr} = 0$ V.

The total interfacial capacitance ($C_{tot}$) was evaluated in terms of electrolyte and electrode capacitances for a given $\sigma$ value, similarly to capacitors in series:

$$1/C_{tot} = 1/C_{IL} + 1/C_{Gr}. \quad (7)$$

The obtained $C_{tot}$ was plotted on the overall potential drop ($U'$) scale that was calculated as the sum of the electrode potential and the electrolyte potential drop for a given $\sigma$ value: $U' = U + \varphi_{Gr}$. Let us note that similar expressions were used to explain experimental results, although in some cases by not evaluating the potential drops, and thus capacitance, within the electrode and electrolyte independently [58].

To investigate changes in the arrangement of interfacial ions at different $T$ and $U$ values, we constructed the number density profiles of the ions in a direction perpendicular to the Gr surface. Furthermore, for better distinguishment between the centers of the interfacial layers, when constructing the number density profiles, the center of the imidazole ring and the location of the boron atom were considered as the positions of the cation and anion, respectively.

## 2.3. Interfacial bilayer model

The layered interfacial structure of IL allows viewing the oppositely charged IL layers as series-connected capacitors [59]. By considering only the first two IL layers to have a significant effect on the $\varphi_{IL}$, then it can be estimated using the interfacial bilayer (IBL) model [60], which defines it as:

$$\varphi_{IL} \approx \frac{l\sigma - \delta\lambda}{\epsilon\epsilon_0}. \quad (8)$$

The $C_{IL}$ [57]:

$$C_{IL} = \frac{d\sigma}{dU} = \frac{d\sigma}{d\varphi_{IL}} \approx \frac{\epsilon\epsilon_0}{l} + \frac{\epsilon\epsilon_0 U - \delta\lambda}{l^2} \cdot \nabla l + \frac{\delta}{l}\nabla\lambda + \frac{\lambda}{l}\nabla\delta, \quad (9)$$

which can be simplified by assuming $l$ = const:

$$C_{IL} \approx \frac{\epsilon\epsilon_0}{l} + \frac{\delta}{l}\nabla\lambda + \frac{\lambda}{l}\nabla\delta, \quad (10)$$

where $\nabla = \frac{d}{dU}$ (partial derivative), $l$ and $\delta$ are the distances between the electrode and the first IL layer and between the first and second IL layers, as illustrated in Fig. 1b. $\lambda$ describes the excess charge of the first IL layer that overscreens $\sigma$. The first term in Eqs. 9–10 is the Helmholtz capacitance:

$$C_\text{H} = \frac{\epsilon\epsilon_0}{l}. \quad (11)$$

Eqs. 9 and 10 are suitable for the semi-qualitative description of various interfacial phenomena [20,23,57,61,62] including the $C_\text{IL}(U)$ dependence.[1] First, the capacitance peak ($C_\text{P}$) is related to the maximal $\nabla\lambda$ value in the vicinity of the PZC. When considering the changes in the positions of the first two that are of IL layers negligible ($\nabla l = 0$ and $\nabla \delta = 0$) and considering the electroneutrality (as $\sigma = \lambda/(\beta-1)$), then Eqs. 8 and 9 give:

$$U_\text{P} \approx \frac{l-\delta(\beta-1)}{\epsilon\epsilon_0}\sigma - \varphi_\text{PZC}, \quad (12)$$

and

$$C_\text{P} \approx \frac{\epsilon\epsilon_0}{l-\delta(\beta-1)} = \frac{l}{l-\delta(\beta-1)}C_\text{H} = bC_\text{H}. \quad (13)$$

The last equation implies that the capacitance peak can be related to the overscreening (in terms of $\delta$ and $\beta$). The formula can be simplified further by taking $\beta = 2$, i.e., assuming that the maximal $\nabla\lambda$ value is achieved when the ion exchange happens *via* desorbing one co-ion and adsorbing one counter-ion while changing the surface charge by one ion-charge per area [64,65]:

$$C_\text{P} \approx \frac{\epsilon\epsilon_0}{l-\delta} = \frac{l}{l-\delta}C_\text{H}. \quad (14)$$

Second, at the saturation potential ($U_\text{S}$), the saturation of the second layer of co-ions take place, when $\nabla l = 0$, $\nabla \delta = 0$, and $\nabla\lambda = 0$ [57]:

$$C_\text{S} \approx \frac{\epsilon\epsilon_0}{l} = C_\text{H}. \quad (15)$$

Below, we discuss the saturation phenomenon in more detail.

Finally, at the monolayer formation potential ($U_\text{M}$), the first layer completely screens the surface charge. Assuming that the screening at that point is Helmholtz-like, the capacitance at $U_\text{M}$ equals [57]:

$$C_\text{M} \approx \frac{\epsilon\epsilon_0}{l+\delta} = \frac{l}{l+\delta}C_\text{H} = aC_\text{H}, \quad (16)$$

---

[1] For electrodes with a limited number of electronic states near the Fermi level (like graphene), the contribution of $C_\text{Q}$ must be added to the model [63].

where, in the crowding regime [8], 0.2 ≤ a ≤ 0.5 with the lowest and highest values being for the close-packing and inverse-square-root scaling [7], respectively. Furthermore, a scaling relation, arising from the charge conservation law, can be used for the estimation of $C_{IL}(U)$ dependence [60]:

$$C_{IL} = a \frac{\epsilon \epsilon_0}{l} \left(\frac{U}{U_M}\right)^{a-1}, (17)$$

where *a* is the scaling exponent. At $U = U_M$ it simplifies to Eq. 16, and, if $a = 0.5$, it follows the inverse-square-root scaling.

In the studied potential range from −6.5 to 6.5 V, the Gr–EMImBF$_4$ interface is in the over-screening regime [18,66], as the monolayer formation potential ($U_M$) for EMIm$^+$ and BF$_4^-$ are −13 V and 20 V, respectively [67]. Accordingly, we focus on the three main variables in Eq. 10 to rationalize the simulated data: Helmholtz capacitance ($C_H$), excess charge density derivative ($\nabla \lambda$), and the layering distance derivative ($\nabla \delta$). $C_H$ is defined by the *packing* of counter-ions, $\nabla \lambda$ is dictated by the EDL *charging* via sorption of ions, and $\nabla \delta$ is determined by ion *layering*. Below we present the results using the named concepts along with three characteristic potentials: (1) $U_{PZC}$ when the EDL is overall neutral, (2) $U_S$ when the second layer is saturated with co-ions (see Fig. 2), and (3) $U_M$ when the first layer is saturated with counter-ions and completely screens the surface charge.

# 3. Results and discussion

## 3.1. The effect of electrode potential on electrolyte capacitance

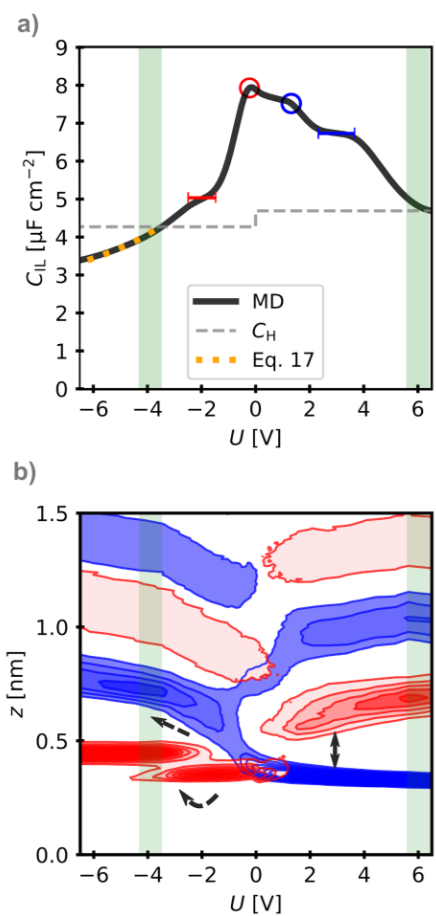

Figure 2. a) Calculated $C_{IL}(U)$ curve for the Gr–EMImBF$_4$ interface simulated at 450 K. The grey and orange lines show the Helmholtz capacitance ($C_H$) and the dependence fitted using Eq. 17. The red and blue markers indicate the critical points and plateaus in the $C_{IL}(U)$ dependency. b) The number density $\rho_N(z, U)$ contour maps of EMIm$^+$ (red area) and BF$_4^-$ (blue area) ions at $T$ = 450 K. The interval of contour lines is equal to $\rho$ bulk, with the first contour starting at 1.25$\rho_{bulk}$ and the last contour representing densities higher than 8$\rho_{bulk}$. Arrows mark the potential regions where (1) the layering of IL (bidirectional arrow), (2) the reorientation of cation (curved arrow), and (3) saturation of the second IL layer (straight arrow) occur. Green regions in both figures show the saturation potentials ($U_S$), where the second IL layer reaches its maximum densities.

Fig. 2 shows the $C_{IL}(U)$ dependence and the number density contour map that reflects the changes in the EDL structure. We regard the $C_{IL}(U)$ neither as a camel or bell-shaped but rather as an overlap of two peaks due to the anion and cation adsorption at positive and negative potentials. The overlap of these peaks leads to a wide maximum at the PZC. The arrows in Fig. 2b points to the relations between the anionic and cationic capacitance peaks and the changes of IL structure. These structural changes can be expressed as the maximal values of $\nabla\delta$, $\nabla\lambda$, and $\nabla l$ in Eq. 10. Using Eq. 14, we estimated $C_P$ near the PZC as 9 µF/cm² and 7 µF/cm² at U = −0.1 V and 0.1 V, respectively. The evaluated $C_P$ values are in relatively good agreement with the calculated $C_{IL}(U)$ curve in Fig. 2a. Therefore, the IBL model allows making crude estimations of the capacitance peaks.

The $C_{IL}$ is independent of $U$ within −2.5–−1.5 V and 2–4 V, indicated by the two plateaus in the $C_{IL}(U)$ dependence (see Fig. 2a). At negative potentials, the density contours indicate the reorientation of EMIm⁺ ions between parallel and perpendicular directions relative to the Gr surface. Some aspects of such reorientation are discussed in Refs. [68–70]. As in previous works [41,43], with the reorientation of cations upon surface charging, $\delta$ remained almost the same. In this work, we note that upon surface charging $l$ effectively increases (see Fig. 2b). Within a specific $U$ range, the first two terms in Eq. 9 balance for the change of $l$. That results in an apparently constant $C_{IL}$ value around −2 V.

As shown in Fig. 2b and Fig. S1, a similar compensating effect occurs around 3 V. The mechanism is slightly different. Here, due to the smaller size of ions, a relatively low counter-ion density is required to overscreen the surface charge. That leaves voids on the electrode surface, which are filled with cations, as shown for the $U$ values 1.5 V (12 µC cm⁻²) and 2.1 V (16 µC cm⁻²) in Fig. S2. Upon further surface charging, the first IL layer becomes denser, and the cations or their alkyl chains can no longer settle in the voids of the first layer. That causes a more significant separation of the two oppositely charged IL layers and the increase of $\nabla\delta$. Meanwhile, the number of co-ions in the second IL layer does not significantly increase (see SI Fig. S1b), indicating the decrease of $\nabla\lambda$ and leading to a constant $C_{IL}$ value in the $U$ range of 2–4 V. In our previous work [57], a similar mechanism resulted in a capacitance peak caused by the BMIm+ reorientation.

Fig. 2b shows that the second layer has its maximal density at −4 V and 6 V. These are the saturation potential ($U_S$) for co-ions in the EDL structure. At absolute potentials larger than $|U_S|$, the capacitance curve follows Eq. 17 [60] with fitted $a$ values of 0.58 (for $U < -4$ V, see Fig. 2a) and 0.81 (for $U > 6$ V). In additional MD simulations (see Appendix 3), we estimated the $C_H$ values of 4.3 μF/cm² (for Gr–EMIm+ interface) and 4.7 μF/cm² (for Gr–BF$_4^-$ interface). Fig. 2a highlights a reasonably precise intersection of vertical lines ($U_S$) and the dashed line ($C_H$) with the calculated $C(U)$ curve. That is expected from the IBL model (Eqs. 9 and 15) assuming $\nabla l = 0$, $\nabla \delta = 0$, and $\nabla \lambda = 0$, because the latter condition expresses the saturation. Moreover, the IBL model enables estimating the combining $U_S$ by combining Eqs. 15 and 17:

$$U_S = U_M \left(\frac{1}{a}\right)^{1/(a-1)}, (18)$$

The evaluated values of −3.5 V and 6.5 V are in an acceptable agreement with the simulation results from Fig. 2, where the saturation is seen around −4 V and 6 V, respectively.

As shown, the IBL model consistently interprets the calculated $C_{IL}(U)$ dependency. Namely, the following $C_{IL}(U)$ features are seen in Figure 2a: (1) $C_{IL} \approx aC_H(U/U_M)^{a-1}$ scaling at U < −4 V (Eqs. 16–18), (2) $C_{IL} \approx C_H$ at the saturation potential (Eq. 15), (3) plateaus around −2 and 3 V, (4) $C_P \approx bC_H$ at the peak potential (Eq. 13–14).

## 3.2. The effect of temperature on the electrolyte capacitance

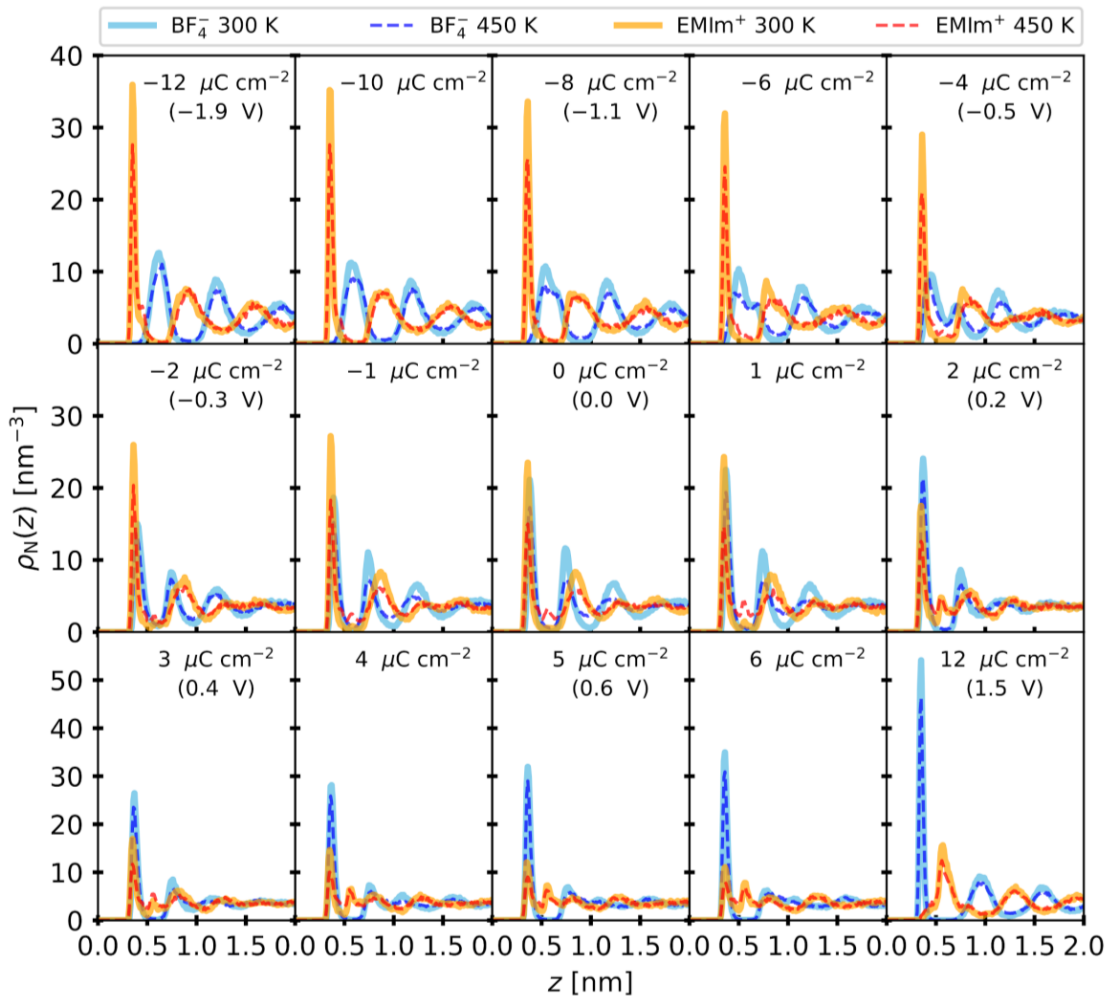

Figure 3. Number density profiles of EMIm$^+$ and BF$_4^-$ ions in the case of the $U$ range from −1.9 to 1.5 V at $T$ values 300 K and 450 K. The potential $U$ values are given for 450 K.

Fig. 3 illustrates the changing of the Gr–EMImBF$_4$ interfacial structure in the $U$ range from −1.9 to 1.5 V. The IL has a layered structure at the interface. The layers contain approximately equal amounts of anions and cations even at the PZC. The formation of a layered structure is caused by the interaction of ions with Gr, which prevents the ions from packing similarly to the bulk IL. As the electrode is charged, co-ions are pushed away from the surface to the second layer. That can be seen in Fig. 3 by comparing the profiles of $U$ values −0.5 and −1.1 V as well as 0.0 and 0.2 V. Despite the electrostatic repulsion between the electrode and co-ions, the first IL layer contains a significant number of co-ions at small |U|, enabled by ion–ion correlation and the voids on the electrode surface that are large enough to accommodate co-ions.

Temperature affects the capacitance foremost near the PZC, as shown in Fig. 4. Within −0.5–2 V, when the $T$ increases, the values of $C_{IL}$ peaks decrease, and the separation between two local capacitance peaks increases. When comparing the number density profiles of $T$ equals 300 K and 450 K in Fig. 3, it can be seen that the increased thermal motion leads to changes in the IL interfacial structure. On the one hand, the position of the number density peaks remains the same. On the other hand, all peaks are smeared due to thermal motion – their heights are shortened, and widths are broadened. Thus, we may assume that $l$ and $\delta$ in the IBL model do not depend on the $T$, while $\beta$ and $\lambda$ do. That means the thermal motion foremost diminishes overscreening and decreases the maximal excess charge density. These structural changes translate into a particular $C(U,T)$ dependence.

According to Eqs. 12 and 13, with the decrease of $\beta$, the peak absolute potential ($U_P$) value increases, while the capacitance peak ($C_P$) value decreases, resulting in $dC/dT < 0$. Fig. 4 illustrates the same trends in the MD simulations. The reader can mark the $|U_P|$ increase (especially on the positive side), and the $C_P$ decreases, decrease with the increase of $T$. Note how the shifting of $U_P$ to higher absolute potentials uncovers regions (for example, around 1–2 V) where $dC/dT > 0$. That leads us to the essential question of choosing the proper reference for the $dC/dT$ comparison. One can choose between arbitrary $U$ or $\sigma$ values and compare differential and integral capacitances, like in Ref. [41]. On top of that, there are fast and slow capacitive processes [71], which analysis requires investigating the impedance spectra. Considering all this, we suggest using characteristic potentials such as $U_M$, $U_S$, or $U_P$. From Eqs. 10 and assuming $l$ and $\delta$ are independent of $U$ and $T$, it is straightforward to express the $C(T)$ dependencies at these characteristic potentials:

$$\frac{dC_M}{dT} \approx 0, \frac{dC_S}{dT} \approx 0, \text{ and } \frac{dC_P}{dT} \approx \frac{\delta}{l}\frac{d\nabla\lambda}{dT} \quad (19)$$

The first potential ($U_M$) [66], is hardly achievable in experiments [72]. Moreover, it is independent of the $T$, as the strength of electrostatic electrode–ion interactions exceed the thermal energy at high $|U|$. The second potential ($U_S$) can possibly be probed with atomic force microscopy and X-ray reflectivity measurements, yet it is independent of T according to Eqs. 15 and 19. The last potential ($U_P$) is most easily identified in experiments, simulations, and theory as a striking $C(U)$ curve feature. Focusing on $U_P$ has the advantage of linking the $C(U,T)$ dependence to the EDL charging mechanism in terms of *packing* ($C_H$), *layering* ($\nabla\delta$), and *charging* ($\nabla\lambda$). Herewith, choosing an arbitrary constant $U$ value may lead to wrong conclusions. For example, it would be ungrounded to relate the positive dependence within 1–2 V from Fig. 4 to the anomalous capacitance concept from the works by Boda *et al.* [30–32]. Therefore, we propose focusing on $dC_P/dT$ at $U_P$ to avoid misinterpretations. In this study, the $\lambda$ and, accordingly, $C_P$ values decreased with temperature resulting in $dC_P/dT < 0$ (see Fig. 4).

Qualitatively the results of our simulations resemble the experiments of Drüschler *et al.* [38] with BMPyrFAP IL at Au(111) electrode. First, they also observed a wide capacitance peak that we interpret as overlapping anionic and cationic peaks. Second, the capacitance peak value decreased with increasing temperature ($dC_P/dT < 0$). Third, similarly to Fig. 4, their anionic peak shifted when increasing $T$. The latter resulted in both $dC/dT < 0$ and $dC/dT > 0$ being observed in different $U$ regions. These complex results described the *fast capacitive process* of restoring electroneutrality upon potential variation. The *slow capacitive process* was characterized by $dC/dT < 0$ at all studied potentials, and it has not been captured in the presented simulations of the Gr–EMImBF$_4$ interface. Experimental results suggest that the slow process may be related to the reorientation of strongly bound ions. Thus, in principle, it could be observed around −3 V for the Gr–EMImBF$_4$ interface if it is studied with considerably longer non-equilibrium MD simulations, which allow investigating the dynamic processes occurring at the interface [68]. Drüschler *et al.* [38] likewise pointed out that the differentiation between the slow and fast processes is essential in understanding the $C(U,T)$ dependence. Due to the lack of such analysis in other experimental works, we refer to their discussion in Ref. [61] and overlook the so-called anomalous dependence ($dC/dT > 0$) [73].

Previous computational works, relying on MD simulations, reported in [41,43,74,75], confirm our observations that the $T$ increase does not induce any significant separation of IL layers nor peak positions changes in the density profiles. Concurrently, the $T$ increase shortens and widens the number density peaks. $C(U)$ curves at different $T$ from Refs. [39–41] display the same features as reported in this work, albeit in a narrower studied potential range. Namely, $dC_P/dT < 0$ near the PZC, the shift of $U_P$ to higher absolute potentials, and ranges with $dC/dT > 0$. Chen *et al.* [40] describe the separation of capacitance peaks as the transformation of $C(U)$ from camel to bell shape within the modified mean-field theory. Shen *et al.* [42] studied the same Gr–EMImBF$_4$ interface with classical DFT and observed similar features of the $C(U,T)$ dependence as in this study: $dC_P/dT < 0$; $U$ ranges with $dC/dT > 0$; and shift of $U_P$ upon heating. These and other authors [76] referred to the idea of ion association that weakens with increasing $T$. It may be speculated that the observed changes in overscreening ($\beta$), saturation ($\lambda$), reorientation and layering ($\delta$) are also related to the electrode–ion and ion–ion correlation affected by $T$. Thus, we find that the proposed IBL model-based explanation of the $U_P$ shift and $dC_P/dT < 0$ is complementary to both theoretical explanations [40] and phenomenological reasoning [39,41] of the $C(U,T)$ dependence.

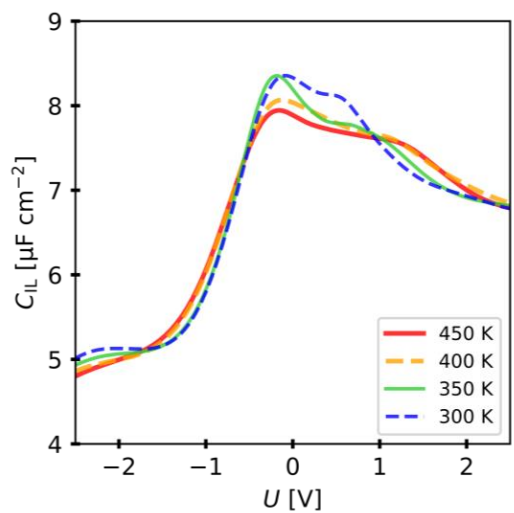

Figure 4. Gr–EMImBF$_4$ interface $C_{IL}(U)$ dependencies for simulated $T$ values in the potential range from −2.5 to 2.5 V. The potential axis ($U$) accounts for the shift due to $\varphi_{PZC}$ of each studied $T$. The average values of the $\varphi_{PZC}$ were 0.17 V, 0.14 V, 0.13 V, and 0.11 V at 300 K, 350 K, 400 K, and 450 K, respectively.

### 3.3. The effect of temperature and quantum capacitance on interfacial capacitance

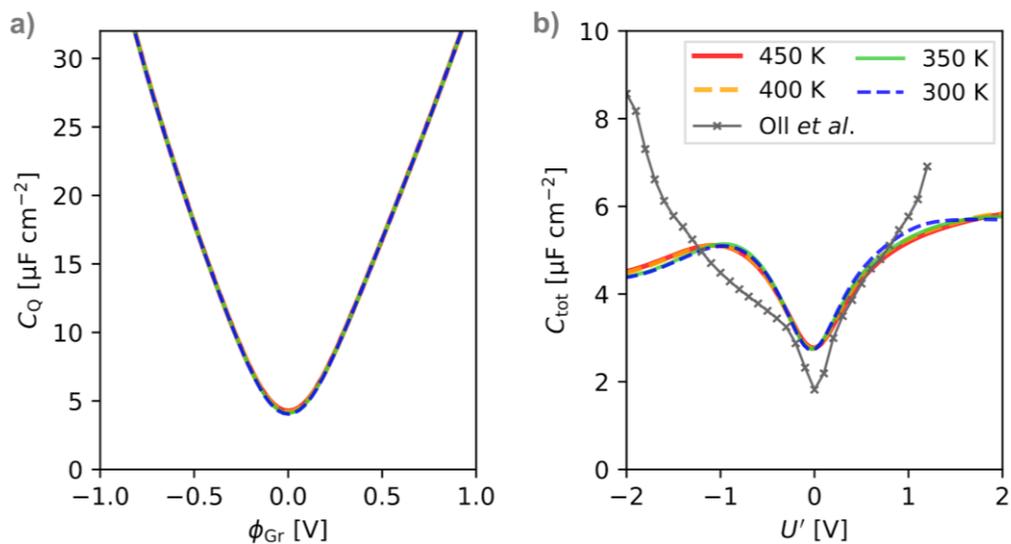

Figure 5. a) $C_Q(\varphi_{Gr})$ dependences for simulated temperatures. b) $C_{tot}(U')$ dependences for simulated temperatures along with the experimental data reported by Oll et al. [16]. The potentials of experimental results by Oll et al. are shifted by the PZC.

A notable potential drop within the Gr is observed due to significant shifting of the Fermi level, caused by the low density of states near the latter. Fig. 5a shows the calculated quantum capacitance ($C_Q$) with a characteristic V shape in agreement with the experimental results reported by Xia *et al.* [77] and Fang *et al.* [78]. Furthermore, Fig. 5 highlights the negligible effect of $T$ on the $C_Q$. When comparing the $C_Q$ and $C_{IL}$ values near the PZC, $C_Q$ is almost two times lower and therefore limits the total interfacial capacitance, shown in Fig. 5b. As the $C_Q$ has a substantial effect on the $C_{tot}$ within the $U'$ range from −1 V to 1 V, where the $C_{tot}(U')$ curve has a minimum in contrast to the $C_{IL}(U)$ maxima. Thus, as the $C_Q$ is only slightly affected by the $T$, the heating of the studied system has a minor effect on $C_{tot}$ in the studied $T$ range.

The impact of $C_Q$ on $C_{tot}$ of Gr–BMImPF$_6$ interface was previously studied with MD simulations in combination with DFT calculations by Paek *et al.* [79], where the limiting effect of $C_Q$ on $C_{tot}$ was also demonstrated. However, the effect was not as drastic, due to lower $C_{IL}(U)$ maxima. For the Gr–EMImBF$_4$ interface, a similar trend was reported by Zhan *et al.* [63], who used a coarse-grained model of EMImBF$_4$. Experimentally, using electrochemical impedance spectroscopy, the Gr–EMImBF$_4$ interface was studied by Oll *et al.* [16], who showed that interfacial $C(U)$ dependence has a minimum near the PZC. As shown in Fig. 5b, within the $U'$ range from −1 V to 1 V, their reported $C(U')$ dependence is in accordance with our simulations. The experimental curve lacks a "shoulder" around −1 V and at higher absolute potentials exceeds our estimated $C(U)$ values. The deviation may be caused by several approximations used in the MD simulations and analysis, such as (1) simplistic electrode model, (2) simplistic electronic polarization model, and (3) poor sampling. The straightforward way of addressing all these issues is in running longer simulations with constant potential methods and polarizable force fields [21,80–83]. That way has been recently opened due to software development, force field derivation, and DFT-based MD reference simulations [25,26,82,84–86]. We proceed with it to obtain models reproducing experimental results.

## 4. Conclusions

In this study, we modeled graphene–ionic liquid (IL = EMImBF$_4$) interfacial structure and capacitance ($C$) dependence on potential ($U$) and temperature ($T$). We have defined three structure-determined potentials ($U_M$, $U_S$, and $U_P$) to interpret the simulated $C(U,T)$ curves using the original interfacial bilayer model. The maximal density of counter-ions of the first layer and co-ions of the second layer is reached at $U_M$ and $U_S$, respectively. The capacitance peak potential ($U_P$) corresponds to a state when the excess charge density derivative ($\nabla\lambda$) is at the maximum. For these potentials, the interfacial bilayer model approximates $C(U,T)$ as:

$$C_P \approx \frac{l}{l-\delta} C_H \text{ and } \frac{dC_P}{dT} \approx \frac{\delta}{l}\frac{d\nabla\lambda}{dT},$$

$$C_S \approx C_H \text{ and } \frac{dC_S}{dT} \approx 0,$$

$$C_M \approx \frac{l}{l+\delta} C_H \text{ and } \frac{dC_M}{dT} \approx 0,$$

where $l$ and $\delta$ are the distances between the electrode and the first IL layer and between the first and second IL layers, and $C_H$ is the Helmholtz capacitance. We suggest using the capacitance peak potential ($U_P$) for analyzing the $C(T)$ dependence. That allows linking structural changes with interfacial properties. For instance, we observed negative dependence ($dC_P/dT < 0$) and showed its correlation to excess charge density ($\lambda$) decrease with increasing $T$.

In essence, the listed equations relate the geometry ($l$, $\delta$) with the capacitance without any empirical parameters. However, the underlying model requires a correction for the electronic density distribution in the electrode material. We have shown that accounting for the contribution of graphene to the total interfacial capacitance ($C_{tot}$) dampens the $T$ influence and leads to an encouraging agreement with the experimental $C(U)$ curve.

# Acknowledgments

This work was supported by the Estonian Research Council grant PSG249 and by the EU through the European Regional Development Fund under project TK141 (2014-2020.4.01.15-0011). The financial support from FCT/MCTES through the Portuguese national funds, project No. UID/QUI/50006/2021 (LAQV@REQUIMTE) is also acknowledged. For providing us with the computational resources, we would like to recognize the Partnership for Advanced Computing in Europe (PRACE), the Distributed European Computing Initiative (DECI), and the HPC Center of the University of Tartu.